\documentclass[journal,tvcgpapersize]{vgtc}   

\graphicspath{{figures/}{pictures/}{images/}{./}} 

\usepackage{times}                     

\usepackage{booktabs}                  
\usepackage{enumitem}

\usepackage{mathptmx}                  
\usepackage{amsmath}

\onlineid{0}

\vgtccategory{Research}

\vgtcinsertpkg
  \renewcommand*{\backrefalt}[4]{}

\makeatletter
\let\quest@origincludegraphics\includegraphics
\renewcommand{\includegraphics}[2][]{%
  \def\quest@graphicsarg{#2}%
  \def\quest@diamondrule{diamondrule}%
  \ifx\quest@graphicsarg\quest@diamondrule
    \rule{0.72\linewidth}{0.4pt}%
  \else
    \quest@origincludegraphics[#1]{#2}%
  \fi
}
\makeatother

\makeatletter
\renewcommand{\ps@plain}{%
  \renewcommand{\@oddhead}{}%
  \renewcommand{\@evenhead}{}%
  \renewcommand{\@oddfoot}{\hfil\textrm{\thepage}\hfil}%
  \renewcommand{\@evenfoot}{\@oddfoot}%
}
\renewcommand{\ps@empty}{%
  \renewcommand{\@oddhead}{}%
  \renewcommand{\@evenhead}{}%
  \renewcommand{\@oddfoot}{\hfil\textrm{\thepage}\hfil}%
  \renewcommand{\@evenfoot}{\@oddfoot}%
}
\makeatother
\manuscriptnote{This work has been submitted to the IEEE for possible publication. Copyright may be transferred without notice, after which this version may no longer be accessible.}

\title{Divergent Perceptuomotor Recalibration in Virtual Reality and Video-Passthrough Mixed Reality on the Same Head-Mounted Display}

\author{Xiaoye Michael Wang\thanks{e-mail: michaelwxy.wang@utoronto.ca}\\ %
        \scriptsize University of Toronto %
\and Grant Monahan\thanks{e-mail: grant.monahan@mail.utoronto.ca}\\ %
     \scriptsize University of Toronto %
\and Timothy N. Welsh\thanks{e-mail: t.welsh@utoronto.ca}\\ %
     \scriptsize University of Toronto} %

\abstract{
    Virtual reality (VR) and video-passthrough mixed reality (MR-VPT) can be delivered on the same headset, but it is unclear whether these two interaction modalities produce comparable perceptuomotor behavior. Although delivery through the same headset controls many display-level characteristics, VR and MR-VPT differ in both their visual-feedback pipelines and the action-relevant visual information available to guide movement, such as whether the surrounding environment and the user's body are synthetically rendered or preserved through passthrough. This study compared visually guided manual pointing in VR and MR-VPT using the same headset. Forty adults were assigned to either a VR or MR-VPT group and completed a pointing task in the physical, unmediated reality (UR) before and after performing the same task in their assigned XR modality. The analyses revealed numerous important insights. First, groups had comparable baseline performance in UR prior to XR exposure. Second, the two modalities produced opposite initial biases during XR exposure: the VR group undershot the target, whereas the MR-VPT group overshot. Third, although both groups reduced error with continued exposure to XR, there were differences in the patterns of adaptation: the VR group adapted more slowly and exhibited stronger distance-dependent undershooting than the MR-VPT group. Finally, after XR exposure, both groups showed significant undershooting aftereffects with the VR group retaining a persistent residual undershoot, suggesting incomplete de-adaptation. Together, the results suggest that perceptuomotor recalibration is shaped not only by display-level constraints but also by the action-relevant visual information available within the interaction environment. These findings highlight the need to validate perceptuomotor interaction techniques separately in VR and MR-VPT, even when they are implemented on the same hardware.
} 


\begin{document}


\firstsection{Introduction}

\maketitle

Extended reality (XR) systems use head-mounted displays (HMDs) that integrate display, sensing, and interaction technologies to present digital content for naturalistic interaction in three-dimensional (3D) space. Here, XR is used as an umbrella term encompassing a spectrum of immersive technologies that blend physical and virtual environments \cite{Laato:2024:HICSS}. In virtual reality (VR), users interact within a fully computer-generated environment that visually replaces the physical world. In mixed reality (MR), digital content is presented in the context of and spatially co-registered with the physical environment to support stable object placement and interaction. There are two common HMD-based MR implementations. Video-passthrough MR (MR-VPT) uses the HMD's outward-facing cameras and sensors to capture the physical environment and render it on the headset's displays, allowing virtual content to be composited onto a camera- and display-mediated view of the real world. In contrast, optical see-through MR (MR-OST) uses a transparent combiner (e.g., waveguide-based optics; \cite{Ding:2023:Elight}) to optically overlay virtual imagery while allowing users to view the physical environment without the mediation of the displays.

The increasing availability of HMDs that support both VR and MR-VPT raises a fundamental question for XR interaction: whether these modes should be expected to produce comparable perceptuomotor behavior when implemented on the same device. This question has become increasingly important with the growing interest in cross-reality systems that support interaction techniques, applications, and transitions across multiple levels of virtuality on a common hardware platform \cite{AudaEtAl:2023:CSUR}. Here, \textit{perceptuomotor behavior} refers to visually guided actions in which perceptual information about the target and the effector is used to guide movement. Because accurate performance depends on a calibrated perception-action mapping, changes in this mapping, either due to changes in the available action-relevant visual information or the user's action capabilities, require \textit{recalibration}. This perceptuomotor recalibration restores the functional scaling between perceptual information and action \cite{BinghamPanMonWilliams:2014:JEPHPP, PanCoatsBingham:2014:JEPHPP}. Relative to natural viewing, HMDs alter the availability and fidelity of action-relevant visual information and, consequently, the perception-action mapping through display-specific optical, temporal, and computational constraints. As a result, accurate interaction in XR requires recalibration as users adjust how visually specified target and effector locations map onto action in the mediated environment. This recalibration is behaviorally expressed as perceptuomotor adaptation, with performance gradually changing over repeated interaction. 

On the one hand, because VR and MR-VPT can share the same opaque display hardware, both modalities expose users to similar display-level constraints relative to behavior in the real, unmediated reality (UR). These constraints include the vergence-accommodation conflict (VAC), in which vergence eye movements are driven by rendered scene depth while accommodation remains fixed at the physical focal distance of the display \cite{BatmazTurkmenSarac:2023:VRST, WangPrenevostTarun:2026:Displays, WangSouthwickRobinson:2024:VR}, as well as motion-to-photon latency introduced by camera-, sensor-, and display-mediated visual feedback loops \cite{FreiwaldKatzakisSteinicke:2018:VRST, WangSabistonWelsh:2026:Latency, Warburton:2023:BRM}. If these shared display-level constraints are the main sources of perturbations that alter the available action-relevant information and, thus, perception-action mapping, then VR and MR-VPT should produce broadly similar patterns of perceptuomotor changes and adaptation.

On the other hand, equivalence in display hardware does not imply equivalence in perceptuomotor behavior. Unlike VR, which fully replaces the visual environment with a synthetic scene, MR-VPT preserves visual access to the physical scene, including real-world structure and the user's own body \cite{ChaurasiaNieuwoudtIchim:2020:CGIT, VaziriLiuAseeri:2017:SAP}. At the same time, MR-VPT is not direct viewing: the physical world is captured by outward-facing cameras and is then reconstructed for the headset displays, so the viewed scene remains mediated by sensing, view synthesis, and display processes \cite{ChaurasiaNieuwoudtIchim:2020:CGIT,KuoPennerMoczydlowski:2023:SIGGRAPH}. Although camera-mediated reconstruction may introduce spatial inconsistencies that are absent in direct viewing, MR-VPT also preserves action-relevant information that is typically absent in fully synthetic VR, including familiar scene structure and body visibility. Consistent with this broader view, recent work found that, relative to VR, MR-VPT produced higher misjudgment rates, longer task completion times, and greater head movement in a series of depth-dependent perceptual and interaction tasks \cite{WestermeierBruebachWienrich:2024:TVCG}. These considerations suggest that, even under similar hardware constraints, VR and MR-VPT may produce different patterns of perceptuomotor adaptation.

XR devices are increasingly positioned as flexible platforms capable of delivering both VR and MR experiences, reducing the need for separate devices and expanding application scenarios \cite{AudaEtAl:2023:CSUR, DeSouzaTartz:2024:FVR}. Whether perceptuomotor adaptation is governed primarily by shared display-level constraints or by the broader action-relevant visual information available within each interaction modality remains unknown. If adaptation differs between VR and MR-VPT despite being delivered on the same HMD, findings obtained in one mode may not generalize to the other. Answering this question is therefore important both for understanding perceptuomotor adaptation in XR and for designing applications that employ or transition between VR and MR-VPT.

The present study addressed this knowledge gap by comparing visually guided manual pointing in VR and MR-VPT delivered on the same HMD. Participants performed visually guided pointing movements in either VR or MR-VPT, as well as in UR before and after the XR task to assess baseline performance and subsequent aftereffects. Specifically, this work demonstrates that:

\begin{enumerate}[label=\arabic*., itemsep=0pt, topsep=0pt]
  \item Despite being delivered on the same HMD, VR and MR-VPT produce opposite initial biases upon entering XR, with undershooting in VR and overshooting in MR-VPT.
  \item Both modalities converge to comparable asymptotic accuracy during XR, but VR stabilizes more slowly than MR-VPT.
  \item Distance-dependent endpoint error is substantially stronger in VR than in MR-VPT and remains stable across XR exposure.
  \item Both modalities produce aftereffects characterized by undershooting in UR, but only the VR group retains a significant residual undershoot by the end of washout.
\end{enumerate}

\section{Related Work}

This section reviews literature relevant to how XR systems shape visually guided action. The section begins with a consideration of display-level perturbations, including VAC and motion-to-photon latency, that can alter perception-action mappings in HMD-based XR. The review then covers action-relevant visual information, including body visibility and scene structure, that may further shape performance across interaction modalities. The next subsection includes an examination of how users recalibrate altered perception-action mappings during XR exposure and how this adaptation transfers back to UR. The final part of the review involves prior comparisons of VR and MR-VPT to identify the remaining gaps in understanding how these modalities differentially shape perceptuomotor adaptation and post-exposure transfer. Together, this work motivates an ecological comparison of how XR modalities shape perceptuomotor adaptation when delivered on the same HMD.

\subsection{Display-Level Perturbations}

Relative to natural viewing, XR systems introduce display-level perturbations that alter the mappings between perception and action. Two prominent sources of these perturbations are VAC and motion-to-photon latency.

To support depth perception, HMD-based XR systems use stereoscopic displays to present slightly different images to the two eyes \cite{LinWoldegiorgis:2015:JSID, WangTroje:2023:VisCognA, WangTroje:2024:JVis}, producing binocular disparity \cite{HowardRogers:1995:BVS, Julesz:1971:FCP} and enabling a vivid 3D perceptual experience comparable to that in UR \cite{Harris:2004:BV}. Binocular viewing involves two coupled oculomotor responses: \textit{vergence}, the inward or outward rotation of the eyes to align the two retinal images, and \textit{accommodation}, the change in lens shape needed to keep the viewed object in focus. Under natural viewing, vergence and accommodation are tightly coupled, such that changes in one influence the other \cite{Eadie:2000:OPTO, Hung:1992:OPTO, HungCiuffredaRosenfield:1996:OPTO}. In most HMD-based XR systems, however, accommodation remains fixed at the display's focal plane while vergence continuously adjusts with rendered scene depth, producing VAC \cite{BanksKimShibata:2013:SPIE, Kramida:2015:TVCG, WannRushtonMonWilliams:1995:VisionRes}. VAC has been shown to impair user interactions by increasing time-to-focus \cite{SpiegelErkelens:2024:JSID}, reducing perceptual image quality \cite{ErkelensMacKenzie:2020:SID}, decreasing movement efficiency \cite{BatmazBarreraMachucaSun:2022:CHI} and accuracy \cite{WangPrenevostTarun:2026:Displays, WangSouthwickRobinson:2024:VR}, and impacting higher cognitive functions \cite{DanielKapoula:2019:SR}. For targeted pointing movements, previous studies have reported increasing undershooting with target distance and proposed a geometrical model of VAC in which the accommodative-vergence response biases vergence inward, thereby disrupting the binocular viewing geometry and perturbing the perception-action mapping used to guide movement \cite{WangPrenevostTarun:2026:Displays, WangSouthwickRobinson:2024:VR}.

Another display-level perturbation is motion-to-photon latency, the temporal delay between a user's physical movement and its visual update on the HMD \cite{Warburton:2023:BRM, ZhaoAllisonVinnikov:2017:VR}. This latency arises from multiple stages of the XR rendering pipeline, including sensor sampling, data transfer, scene rendering, and display refresh. Previous work has shown that latency degrades visually guided interaction by reducing movement accuracy and efficiency while prolonging movement time \cite{FreiwaldKatzakisSteinicke:2018:VRST, Hoyet:2019:AIVR, Waltemate:2016:VRST, WangSabistonWelsh:2026:Latency, YangYueGao:2025:TVCG}. In the context of a manual pointing task, latency causes the displayed hand to lag behind the physical hand, creating a discrepancy between the hand's proprioceptively specified physical position and its visually specified virtual position. Accurate performance therefore requires users to recalibrate this altered perception-action mapping. Recent work suggests that this recalibration reflects at least two complementary processes: small delays primarily increase visuo-proprioceptive misalignment and can produce endpoint overshooting, whereas larger delays encourage strategic slowing to compensate for delayed visual feedback, prolonging movement time \cite{WangSabistonWelsh:2026:Latency}. Together, VAC and motion-to-photon latency illustrate how shared display hardware perturbs perception-action mappings and motivate the expectation that VR and MR-VPT should produce broadly similar patterns of perceptuomotor adaptation if these shared display-level constraints govern recalibration.

\subsection{Action-Relevant Visual Information}

Display-level perturbations are not the only factors that can shape visually guided action in XR. Perceptuomotor behavior also depends on the action-relevant visual information available in the surrounding environment, including information about the body, the effector, 3D objects, and the spatial structure of the scene. In VR, this information must be rendered synthetically and is often simplified for practical interaction. Because commercial HMDs typically track only the user's head and hands, many VR applications represent users using floating virtual hands or controllers rather than a fully tracked body \cite{JiangStreliQiu:2022:ECCV, OcampoGonzalezTeather:2025:ISMAR}. Consequently, the action-relevant visual information available for guiding movement may vary substantially depending on the virtual representation. In MR-VPT, by contrast, users retain camera-mediated visual access to the physical environment and their own body (or at least what is in view of the camera), while still viewing that information through a computational reconstruction pipeline.

Visual access to the body is especially relevant for visually guided action because the body provides spatial information about scale, reachability, and relative position of the effector. Prior work has shown that providing a virtual self-avatar can improve egocentric distance judgments in immersive virtual environments \cite{RiesInterranteKaeding:2008:VRST, MohlerCreemRegehrThompson:2010:Presence, PhillipsRiesKaeding:2010:VR}. These findings suggest that body representation provides action-relevant information for scaling the spatial layout of the environment in relation to the body, thereby improving spatial judgments in HMD-mediated environments. More recent work on video passthrough embodiment extends this idea by integrating a real-time video representation of the user's physical body into VR. Although video passthrough embodiment depends on accurate depth sensing and body segmentation, and can therefore introduce visual artifacts, it has nevertheless been shown to enhance presence and embodiment without degrading task performance, cognitive load, or simulator sickness \cite{WaldowFuhrmannRoth:2026:TVCG, WaldowKleinbeckFuhrmann:2025:TVCG}. Thus, visual access to the body may provide action-relevant information that is absent or reduced in bodyless VR, even when users still receive visual feedback from a rendered hand or controller.

In addition to information about the body, the surrounding scene also provides information that can influence distance perception and reaching. Environmental context can affect perceived distance even when standard depth-informative variables are held constant \cite{WittStefanucciRiener:2007:Perception}. In reaching tasks, allocentric information from surrounding objects is also used to encode target locations, including locations in depth \cite{KlinghammerSchutzBlohm:2016:VR, FiehlerWolfKlinghammer:2014:FrontHumNeurosci}. These findings are relevant to XR because VR and MR-VPT differ substantially in the availability and fidelity of surrounding scene information. A fully synthetic VR scene may provide fewer familiar objects, textures, and body-based spatial references than the physical scene available through MR-VPT, whereas MR-VPT may preserve these contextual sources of information even though they remain camera- and display-mediated.

At the same time, passthrough viewing is not equivalent to direct viewing. In MR-VPT, the physical environment is captured by outward-facing cameras and sensors, and reconstructed for the user's eyes through view synthesis and display processes \cite{ChaurasiaNieuwoudtIchim:2020:CGIT, KuoPennerMoczydlowski:2023:SIGGRAPH}. Because passthrough cameras are not co-located with the user's eyes, these systems must reconstruct perspective-correct views, and the resulting image may depend on camera placement, depth estimation, reprojection, and rendering algorithms. As a result, MR-VPT can preserve visual access to the physical environment while also introducing spatial inconsistencies or artifacts that are absent in direct viewing. Consistent with this possibility, recent work comparing VR and MR-VPT on the same HMD found that MR-VPT produced higher misjudgment rates, longer task completion times, and greater head movement than VR, despite participants often preferring the video passthrough condition \cite{WestermeierBruebachWienrich:2024:TVCG}. Together, this literature suggests that VR and MR-VPT should not be treated simply as different visual presentations of the same display hardware. The different modalities provide different action-relevant information for guiding movement, which may shape perception-action mapping, recalibration, and transfer.

\subsection{Perceptuomotor Adaptation}

Despite the perceptuomotor perturbations introduced by XR systems and their immediate consequences for visually guided action, previous studies have shown that users can recalibrate and regain accurate performance \cite{KohmBabuPagano:2022:TVCG, WrightCreemRegehrWarren:2014}. This finding reflects the perceptuomotor system's ability to recalibrate perception-action mappings when sensory feedback or action-relevant visual information is altered \cite{Bock:2005:EBR, BranddeOliveira:2017:HMS, PanCoatsBingham:2014:JEPHPP}. As a result, the initial XR exposure commonly degrades performance (e.g., slower, less accurate movements; \cite{WangNitscheResch:2025:BBR}), yet performance improves with experience \cite{KohmBabuPagano:2022:TVCG}. Importantly, recalibration acquired during XR may have short- and longer-term consequences for performing similar movements in UR, where adaptation that improves performance in XR can produce negative transfer after returning to the physical environment \cite{WangSouthwickRobinson:2024:SR}.

In a previous study \cite{WangSouthwickRobinson:2024:SR}, participants performed targeted manual pointing movements in UR, then in either VR or MR-OST, and finally again in UR. During the XR phase, VR participants undershot relative to the UR baseline, whereas MR-OST participants overshot. Performance in the UR post-test mirrored these XR biases: VR participants initially continued to undershoot before gradually returning to baseline, whereas MR-OST participants continued to overshoot but returned to baseline more rapidly. One possible explanation for these aftereffects is that prolonged exposure to a fixed accommodative demand can temporarily recalibrate vergence-accommodation coupling, causing the hypothesized inward vergence bias proposed in \cite{WangPrenevostTarun:2026:Displays, WangSouthwickRobinson:2024:VR} to persist after HMD removal and produce residual depth-scaling errors until de-adaptation occurs \cite{NeveuRoumesPhilippe:2016:IOVS, PaulusStraubeEggert:2017:JN, YegoGilsonBaraas:2025:Displays}. Together, these findings suggest that perceptuomotor adaptation provides a sensitive measure of how users recalibrate altered perception-action mappings, while aftereffects and de-adaptation trajectories following XR exposure may provide additional insight into whether this recalibration differs between VR and MR-VPT.

\subsection{VR vs. MR-VPT}

Previous studies have compared depth perception and targeted movements between VR and MR-OST \cite{CidotaCliffordLukosch:2016:ISMARAdjunct, PingLiuWeng:2019:VR, WangNitscheResch:2025:BBR}. However, such comparisons often involve different display architectures, making it difficult to determine whether observed differences reflect XR modality, hardware-specific constraints, or both. By contrast, VR and MR-VPT can be delivered on the same opaque HMD, allowing modality-specific visual information to be compared while holding display hardware constant. Only recently have studies begun to compare VR and MR-VPT under this shared-hardware paradigm. Westermeier et al. \cite{WestermeierBruebachWienrich:2024:TVCG} compared VR and MR-VPT on the same HMD using a battery of depth-dependent tasks, including verbal report, bisection, alignment, and a fine motor hotwire task. Across tasks, MR-VPT produced higher misjudgment rates than VR and, in some tasks, resulted in longer completion times, greater head movement, and poorer task performance. These findings suggest that perceptuomotor performance may depend not only on shared display hardware but also on the broader action-relevant visual information available for guiding action.

Despite these findings, existing comparisons of VR and MR-VPT have primarily characterized perceptual judgments and overall task performance (e.g., \cite{WestermeierBruebachWienrich:2024:TVCG}). It therefore remains unclear whether VR and MR-VPT also differ in how users recalibrate altered perception-action mappings through prolonged exposure, how adaptation unfolds over time, and whether any modality-specific recalibration transfers differently back to UR. Because perceptuomotor adaptation reflects the behavioral consequences of recalibrating altered perception-action mappings, examining adaptation and transfer provides a more direct means of distinguishing whether recalibration is governed primarily by shared display-level perturbations or by the broader action-relevant visual information available within each interaction modality.

\subsection{Research Questions and Hypotheses}

Taken together, the literature reviewed above motivates two competing predictions regarding perceptuomotor adaptation in XR. If adaptation is governed primarily by shared display-level perturbations, then VR and MR-VPT should produce broadly similar patterns of error, adaptation, and de-adaptation upon returning to UR because both are delivered through the same display hardware. Conversely, if these processes also depend on the broader action-relevant visual information available within the interaction environment, then VR and MR-VPT should exhibit different patterns of perceptuomotor adaptation despite their shared hardware. To distinguish between these alternatives, the present study compared visually guided manual pointing in VR and MR-VPT, with performance in UR assessed before and after XR exposure. Based on the expectation that action-relevant visual information contributes to perceptuomotor adaptation beyond shared display-level constraints, the current study tested the following hypotheses: 

\begin{itemize}[itemsep=0pt, topsep=0pt]
    \item[$H_1$] VR and MR-VPT will produce different initial endpoint biases upon entering XR, despite sharing the same HMD.
    \item[$H_2$] Perceptuomotor adaptation will unfold differently across modalities, with MR-VPT exhibiting faster adaptation than VR.
    \item[$H_3$] Distance-dependent endpoint error will be stronger in VR than in MR-VPT, indicating stronger depth-related error scaling.
    \item[$H_4$] Both modalities will produce aftereffects upon returning to UR, but VR will exhibit larger and more persistent residual error than MR-VPT.
\end{itemize}

\section{User Study}

\begin{figure*}[t]
  \centering
  \includegraphics[width=\textwidth]{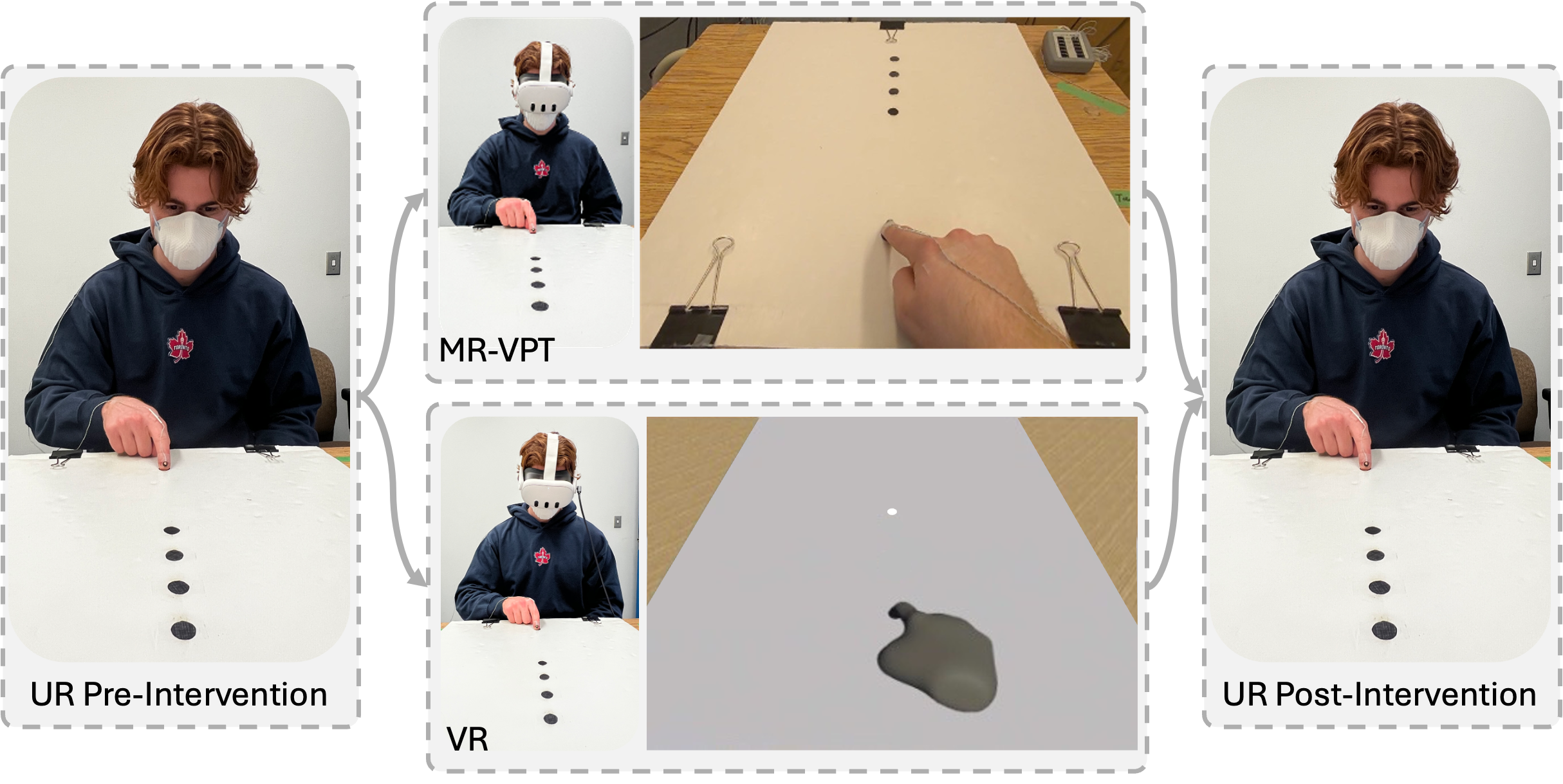}
  \caption{Overview of the same-headset comparison. Participants completed targeted pointing movements in either video-passthrough mixed reality (MR-VPT) or immersive virtual reality (VR) on the same Meta Quest 3 headset. Pointing in unmediated reality (UR) before and after XR exposure was used to assess baseline performance and aftereffects. The first-person panels illustrate the visual information available in each XR modality.}
  \label{fig:study-overview}
\end{figure*}

\subsection{Participants}

Forty right-handed adults between 18 and 31 years of age (mean = 22 years, SD = 2.94) participated in this study, including 20 in the VR group (11 females and 9 males) and 20 (10 females and 10 males) in the MR-VPT group. Data from one participant in the MR-VPT group were excluded from analysis due to motion capture calibration errors, resulting in a final sample of 19 participants in MR. All participants reported normal or corrected-to-normal vision with no known neurological or motor impairments. All procedures were approved by the University of Toronto's Research Ethics Boards. Participants provided full, written informed consent prior to their participation. All participants were naïve to the experimental hypotheses and had minimal prior VR experience.

\subsection{Stimuli and Apparatus}

The experiment used a Meta Quest 3 HMD running at a 90 Hz refresh rate with a resolution of 2064 $\times$ 2208 pixels per eye and a 110° horizontal and 96° vertical field of view. For VR, the HMD was powered by a custom desktop computer with an Intel Core i7 CPU, 32 GB RAM, and an NVIDIA RTX 3080 GPU, connected via Meta Quest Link through a USB Type-C 3.0 cable. 

The general experimental setup was identical in UR, MR, and VR, which consisted of a table with a circular home position and four circular targets (2 cm in diameter) placed at 20, 25, 30, and 35 cm from the home position along the depth axis. Participants sat in front of a physical table. In UR and MR, the home and targets were depicted on a piece of foam board on the physical table. In VR, the virtual environment was built in Unity and mirrored the physical space with a virtual table that shared the same location and height as the physical table to provide haptic feedback. Targets appeared directly in front of the participants, aligned with the midline. The experimental procedures for UR and MR-VPT were controlled using PsychoPy \cite{Peirce:2019:BRM}, whereas those for VR were controlled using bmlTUX \cite{BebkoTroje:2020:iPerception}.

Movements were captured using an opto-electric motion capture system (Optotrak, Waterloo, Canada) via a single infrared-emitting diode (IRED) with a 250 Hz sampling frequency. The IRED was fixed on the participants' right index fingertip using medical tape. In UR and MR, participants could see their physical hand and use it to directly point to the targets. In VR, a 3D virtual hand, formed in a static pointing pose (Figure \ref{fig:study-overview}), was used to provide visual feedback. The virtual hand was animated in real-time using motion capture data from the IRED, transmitted to Unity via a custom Python-Unity interface over a UDP socket. A custom calibration procedure was performed to align the spatial reference frame of the motion capture system to that of VR. In this procedure, the HMD's passthrough mode was programmatically enabled to allow participants to see both the physical and virtual environments. Then, participants sequentially pointed at three non-collinear points in the virtual environment using their physical hand. Fingertip and virtual point positions were used to compute a transformation matrix to convert the kinematic data from the physical reference frame to the virtual reference frame, and the converted kinematic data were used to animate the virtual hand in VR. Using the method described by Warburton et al. \cite{Warburton:2023:BRM}, this system's overall motion-to-photon latency was measured to be around 62 ms (SD = 8 ms). 

\subsection{Design Rationale}

The experiment was designed to compare two realistic XR modalities implemented on the same commercial HMD rather than to isolate a single perceptual variable. Accordingly, several modality-level factors differed together. In MR-VPT, participants viewed the physical table, physical targets, and their physical hand through the headset's passthrough system. In VR, participants viewed a spatially aligned synthetic table and virtual targets, and fingertip feedback was provided by a tracked virtual hand without a full-body avatar. These choices reflect common differences between video-passthrough and VR interaction. The audio cueing used in UR and MR-VPT allowed participants to select among physical targets without adding virtual target overlays, whereas VR used visual target presentation because the targets themselves were virtual. Thus, Modality in this design refers to the implemented perceptuomotor environment as a whole, including scene construction, body representation, target presentation, and display-mediated viewing.

\subsection{Procedures and Design}

\begin{figure}[tb]
  \centering
  \includegraphics[width=\columnwidth]{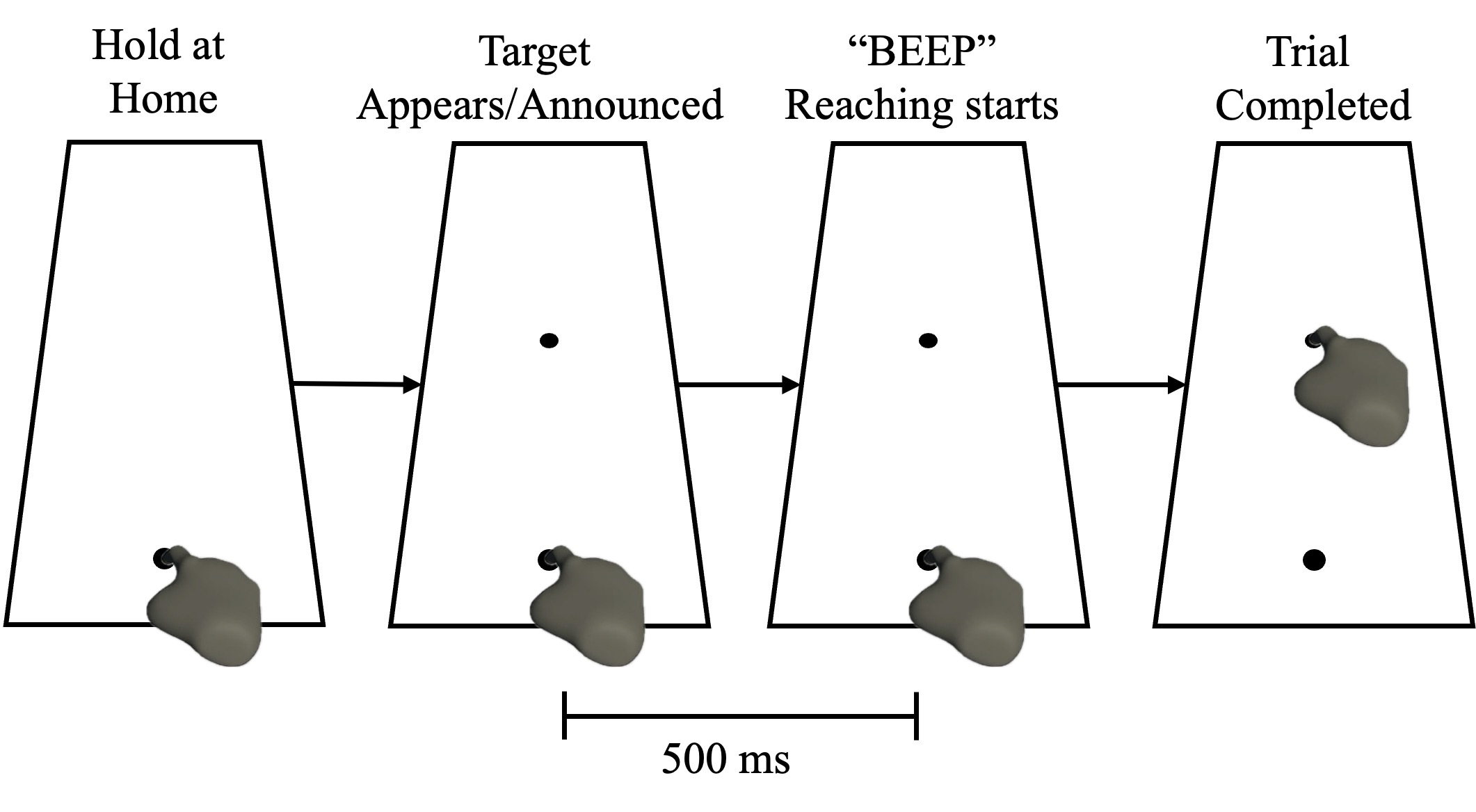}
  \caption{Schematic of a standard trial. Participants first held their right index finger at the home position. Targets were visually present in both XR modalities. In VR, the designated target was shown individually, whereas in UR and MR-VPT, an audio cue indicated which of the four visible targets to point to. After 500 ms, a beep signaled movement initiation. Participants were instructed to point to the target as accurately and as quickly as possible following the beep and hold their finger at the final position until the trial was completed.}
  \label{fig:procedures}
\end{figure}

Figure \ref{fig:procedures} shows the flow of a single trial. At the start of each trial, participants placed their right index finger at the home position, which served as the circular start location for all pointing movements. Four Target Distances were used: 20, 25, 30, and 35 cm from the home position. In UR and MR, all four targets remained visually present and a pre-recorded audio cue specified the target for that trial. After 500 ms, a beeping sound signaled participants to point to the designated target as quickly and accurately as possible. VR followed the same timing, except that only the designated target was presented visually. The target appeared for 500 ms before the beep and remained visible during the movement. Each block consisted of eight trials, comprising two repetitions of each of the four target distances presented in counterbalanced order.

Figure \ref{fig:study-overview} provides an overview of the experimental phases and modality conditions. The study used a between-subjects design for Modality, where participants were assigned to either the VR or MR-VPT group. For both modalities, participants first completed a UR Pre-Intervention phase in the physical environment. This phase consisted of 8 blocks of 8 trials for a total of 64 trials [4 Target Distances $\times$ 2 Repetitions $\times$ 8 Blocks]. Subsequently, participants put on the HMD and completed the XR phase. For the MR-VPT group, participants viewed the physical environment using the HMD's video-passthrough mode, whereas the VR group viewed an immersive virtual environment. In the VR condition, participants completed a calibration procedure to spatially align the virtual and physical environments before the trials. The XR phase consisted of 16 blocks of 8 trials for a total of 128 trials [4 Target Distances $\times$ 2 Repetitions $\times$ 16 Blocks]. Immediately after the XR phase, participants removed the HMD and began the UR Post-Intervention phase in the physical environment with the same setup as the Pre-Intervention phase. This phase also consisted of 16 blocks of 8 trials for a total of 128 trials. The additional blocks in the Post-Intervention phase were included to capture the de-adaptation process. In total, each session consisted of 320 trials across three phases and lasted approximately 1 hour.

\subsection{Data Reduction}

Raw kinematic data were processed in Python (version 3.14.2) using \texttt{TAT-HUM}, a human movement analysis toolkit \cite{WangWelsh:2024:BRM}. Missing data due to marker occlusion were imputed using linear interpolation when they contained no more than 15 consecutive samples, corresponding to 60 ms at 250 Hz. Trials with missing data segments that span more than 15 consecutive samples were discarded. 

Raw displacement data were filtered using a low-pass Butterworth filter with a 10 Hz cutoff frequency. A 3-point finite difference method was used to derive movement velocity and acceleration, smoothed using the same Butterworth filter. Movement onset and termination were determined from the resultant velocity, computed as the Euclidean norm of velocity in the primary and secondary movement directions (depth and height, respectively), using a 100 mm/s threshold. Movement time (MT) was defined as the amount of time between movement initiation and termination. Endpoint error (EE) was defined as the signed deviation between the movement endpoint and the target in depth, with positive values indicating overshoot and negative values indicating undershoot. To retain trial-level variability, raw EE and MT from each trial was used in the subsequent statistical analysis. Variable error (VE) is the standard deviation of EE across repetitions within each unique condition. Although VE was analyzed, it did not reveal any theoretically meaningful effects beyond those already captured by EE and MT, and is therefore not reported here.

\subsection{Statistical Analysis}

All analyses were conducted at the trial level using mixed-effects models in R (version 4.4.3). Across models, the primary predictors were Modality (VR vs. MR), Block, and Target Distance, with participant included as a random effect. A phase refers to one of the three experimental periods (UR Pre-Intervention, XR, and UR Post-Intervention), whereas a block refers to a repeated set of eight trials within a phase. Throughout the paper, adaptation refers to the observed changes in performance across XR blocks, whereas recalibration refers to the inferred adjustment process of the underlying perception-action mapping. Accordingly, the statistical analyses quantify adaptation trajectories, while recalibration is interpreted as the process inferred from those trajectories. Target Distance was mean-centered across the full dataset and expressed in 5 cm units, and Block was indexed from 0. The MR-VPT group served as the reference level throughout. 

The statistical analysis was organized into three parts corresponding to the three experimental phases and the study hypotheses. First, UR Pre-Intervention data were analyzed to establish baseline comparability between the VR and MR-VPT groups before XR exposure. Second, XR data were analyzed to characterize the initial endpoint bias, adaptation process, and distance-dependent EE during XR exposure ($H_1$--$H_3$). Third, UR Post-Intervention data were analyzed to characterize post-exposure aftereffects, de-adaptation, and distance-dependent EE following HMD removal ($H_4$). Because the XR and UR Post-Intervention analyses focused on changes relative to each participant's pre-exposure performance, EE and MT in these phases were baseline-corrected relative to each participant's mean UR Pre-Intervention performance. Specifically, $EE_{bc} = EE - \overline{EE}_{\mathrm{UR\ pre}}$ and $MT_{bc} = MT - \overline{MT}_{\mathrm{UR\ pre}}$. 

\smallskip

\noindent\textbf{UR Pre-Intervention Phase.} Before evaluating the study hypotheses, EE and MT were analyzed to establish baseline comparability between the VR and MR-VPT groups and verify that any subsequent differences could be attributed to XR exposure rather than pre-existing group differences. EE and MT were analyzed with a linear mixed-effects model (LMM) fitted using restricted maximum likelihood, with a BOBYQA optimizer via the \texttt{afex} package \cite{SingmannBolkerWestfall:2015:afex}, which wraps \texttt{lme4} \cite{bates2015lme4}. The model included the main effects of Modality, Block, and Target Distance, as well as all two-way interactions among them, with random intercepts and by-participant slopes for Block. Type~III $F$-tests used the Satterthwaite approximation for denominator degrees of freedom. Effect sizes were reported as partial $\eta^2$ ($\eta_p^2$), computed for fixed effects using the \texttt{effectsize} package \cite{BenShacharLudeckeMakowski:2020:JOSS}. Estimated marginal slopes and pairwise contrasts were obtained with \texttt{emmeans} \cite{lenth2019emmeans} using the Kenward--Roger method.

\smallskip

\noindent\textbf{EE for XR and UR Post-Intervention Phases.} 
To evaluate hypotheses $H_1$--$H_4$, $EE_{bc}$ was modeled separately for the XR and UR Post-Intervention phases using a nonlinear mixed-effects model (NLMM) fitted using maximum likelihood with the \texttt{nlme} package \cite{pinheiro2017nlme}. Both phases used the same exponential-to-asymptote function with an additive linear component for Target Distance. 

\begin{equation}
  EE_{bc}(t) = A + (EE_0 - A)\exp\!\left(-t/\tau\right) + \beta_1\, d + \beta_2\, t\, d
  \label{eq:nlmm}
\end{equation}

The NLMM estimates five quantities that are reported throughout the Results, including the initial bias or aftereffect ($EE_0$), adaptation (or de-adaptation) rate ($\tau$), adapted state ($A$), and distance-dependent EE ($\beta_1$ and $\beta_2$). Since Block was indexed from 0, $EE_0$ denotes the estimated $EE_{bc}$ at the start of the modeled phase for the mean target distance, which quantifies the initial bias upon entering XR (XR phase) or the initial aftereffect upon HMD removal (Post-Intervention phase). $A$ is the long-run asymptote representing the final adapted state reached after extended exposure (XR phase) or the washout period (Post-Intervention phase). $\tau$ is the time constant governing the adaptation (or de-adaptation) rate, expressed in blocks. $d$ is Target Distance, $\beta_1$ captures the main effect of Target Distance on $EE_{bc}$, and $\beta_2$ captures whether the distance-dependent EE changes across blocks. The time constant was log-parameterized by estimating $\ell_\tau = \log(\tau)$ and defining $\tau = \exp(\ell_\tau)$, which enforces $\tau > 0$. By-participant random intercepts were included on $A$ and $EE_0$.

\smallskip

\noindent\textit{Adaptation and de-adaptation parameters.} 
The analysis first evaluated differences in the initial bias or aftereffect ($EE_0$) to test $H_1$ and $H_4$, adaptation or de-adaptation dynamics ($\tau$) to test $H_2$ and $H_4$, and the final state ($A$) to evaluate the persistent residual error prediction in $H_4$. These parameters were modeled as modality-dependent fixed effects. By contrast, $\beta_1$ and $\beta_2$ were initially constrained to be shared across modalities. Modality differences in the nonlinear adaptation parameters were tested with likelihood ratio tests (LRTs) by comparing the base model with nested models in which one parameter ($EE_0$, $A$, or $\tau$) was constrained to be identical across modalities. A significant LRT indicated that relaxing the constraint (i.e., estimating separate modality-specific values) improved model fit, implying that the corresponding parameter differed between VR and MR.

\smallskip

\noindent\textit{Distance-dependent endpoint error.} 
Previous studies proposed that VAC manifests behaviorally as a distance-dependent increase in undershooting, such that EE becomes progressively more negative as target distance increases \cite{WangSouthwickRobinson:2024:VR, WangPrenevostTarun:2026:Displays}. Consequently, to evaluate $H_3$, distance-dependent EE was evaluated by testing whether the distance-related parameters ($\beta_1$ and $\beta_2$) differed between modalities. The base NLMM was compared with nested models allowing these parameters to vary by Modality. In the XR phase, allowing $\beta_1$ to vary by modality significantly improved model fit. However, simultaneously estimating modality-specific distance slopes and modality-specific adaptation parameters introduced strong dependence between the adaptation and distance-scaling parameters, rendering the VR time constant numerically unidentifiable (estimated $\tau > 80$ blocks with a very wide confidence interval). The adaptation parameters ($EE_0$, $A$, and $\tau$) were therefore reported from the base model, whereas modality-specific distance slopes were reported from the expanded model. Because the expanded model was used solely to evaluate distance-dependent EE, this approach did not affect estimation of the primary adaptation parameters. In the Post-Intervention phase, the expanded NLMM was numerically singular. Consequently, distance-dependent washout was evaluated using a supplementary LMM with Modality, Target Distance, a second-order polynomial in Block, and their interactions, with Modality-specific distance slopes estimated using \texttt{emmeans}.

\smallskip

\noindent\textbf{MT for XR and UR Post-Intervention Phases.} 
To complement the EE analyses, MT trajectories during the XR and UR Post-Intervention phases were evaluated using a quadratic LMM. Unlike EE, which exhibited a clear exponential approach toward an adapted state, movement time showed comparatively modest, non-monotonic changes over practice. Consequently, MT was modeled using a quadratic LMM to flexibly characterize its trajectory rather than imposing an exponential adaptation model. Accordingly, $MT_{bc}$ was modeled using a quadratic LMM including mean-centered Block ($block_c$) and $block_c^2$ (to reduce collinearity), together with Modality, Target Distance, and their interactions. Random intercepts and by-participant slopes for $block_c$ were included. The linear and quadratic Block terms jointly characterized the adaptation trajectory of MT, whereas the Modality $\times$ $block_c^2$ interaction tested whether the trajectory differed between VR and MR.

\section{Results}

\begin{figure*}[t]
  \centering
  \includegraphics[width=\textwidth]{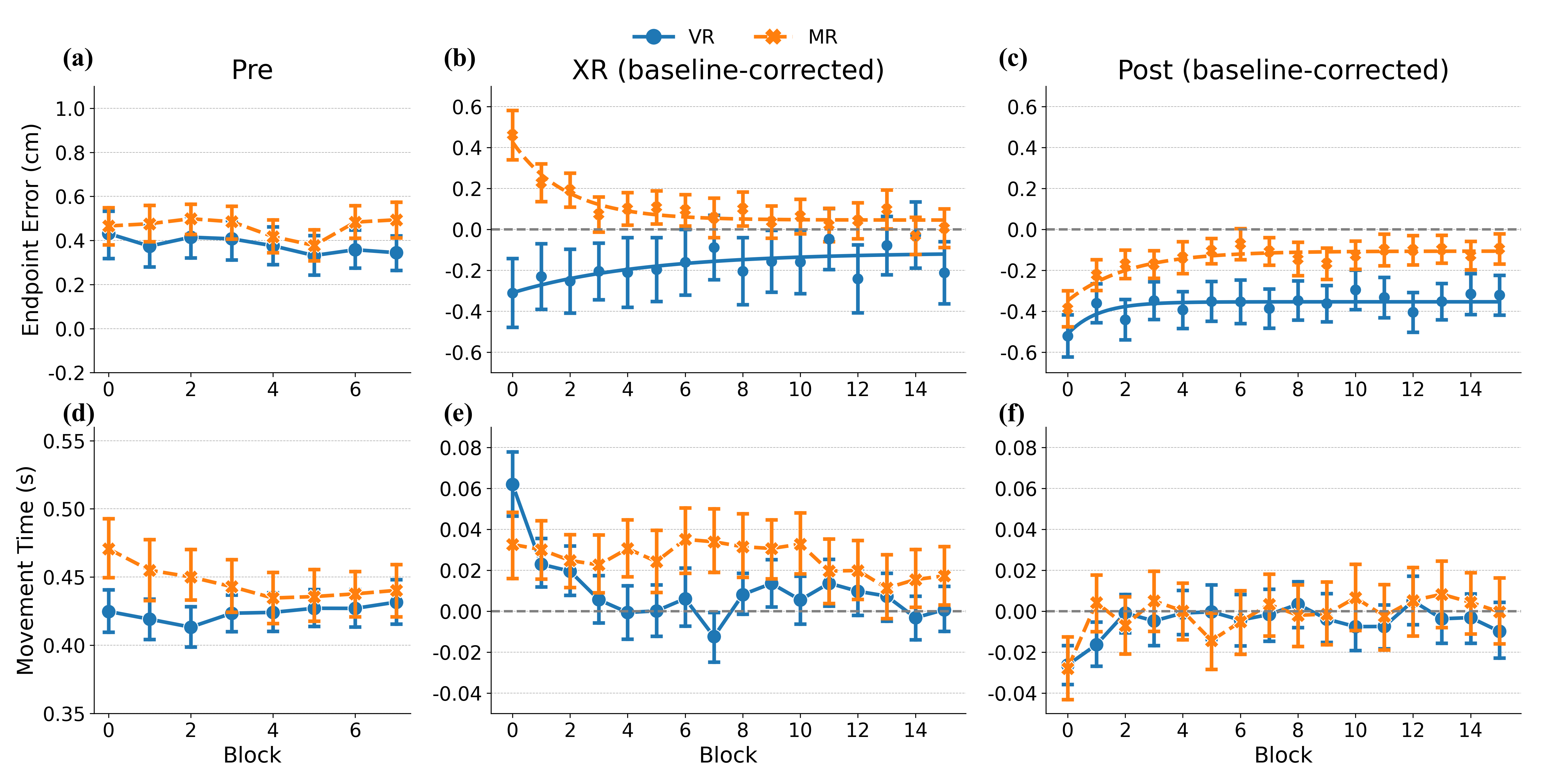}
  \caption{Endpoint error (EE; top row) and movement time (MT; bottom row) for the VR (blue circles) and MR-VPT (orange crosses) groups across the UR Pre-Intervention (left column), XR (middle column), and UR Post-Intervention (right column) phases. \textbf{(a)} Mean signed EE during the Pre-Intervention phase across blocks 0--7. \textbf{(b)} Mean baseline-corrected EE ($EE_{bc}$) during the XR phase across blocks 0--15. Solid and dashed curves show nonlinear mixed-effects exponential-to-asymptote fits evaluated at the mean target distance. \textbf{(c)} Mean $EE_{bc}$ during the Post-Intervention phase across blocks 0--15. Solid and dashed curves show nonlinear mixed-effects exponential-to-asymptote fits evaluated at the mean target distance. \textbf{(d)} Mean MT during the Pre-Intervention phase across blocks 0--7. \textbf{(e)} Mean baseline-corrected MT ($MT_{bc}$) during the XR phase across blocks 0--15. \textbf{(f)} Mean $MT_{bc}$ during the Post-Intervention phase across blocks 0--15. Baseline-corrected values were computed by subtracting each participant's mean Pre-Intervention value. Error bars represent 95\% confidence intervals.}
  \label{fig:ee_mt}
\end{figure*}

\begin{figure}[tb]
  \centering
  \includegraphics[width=.8\columnwidth]{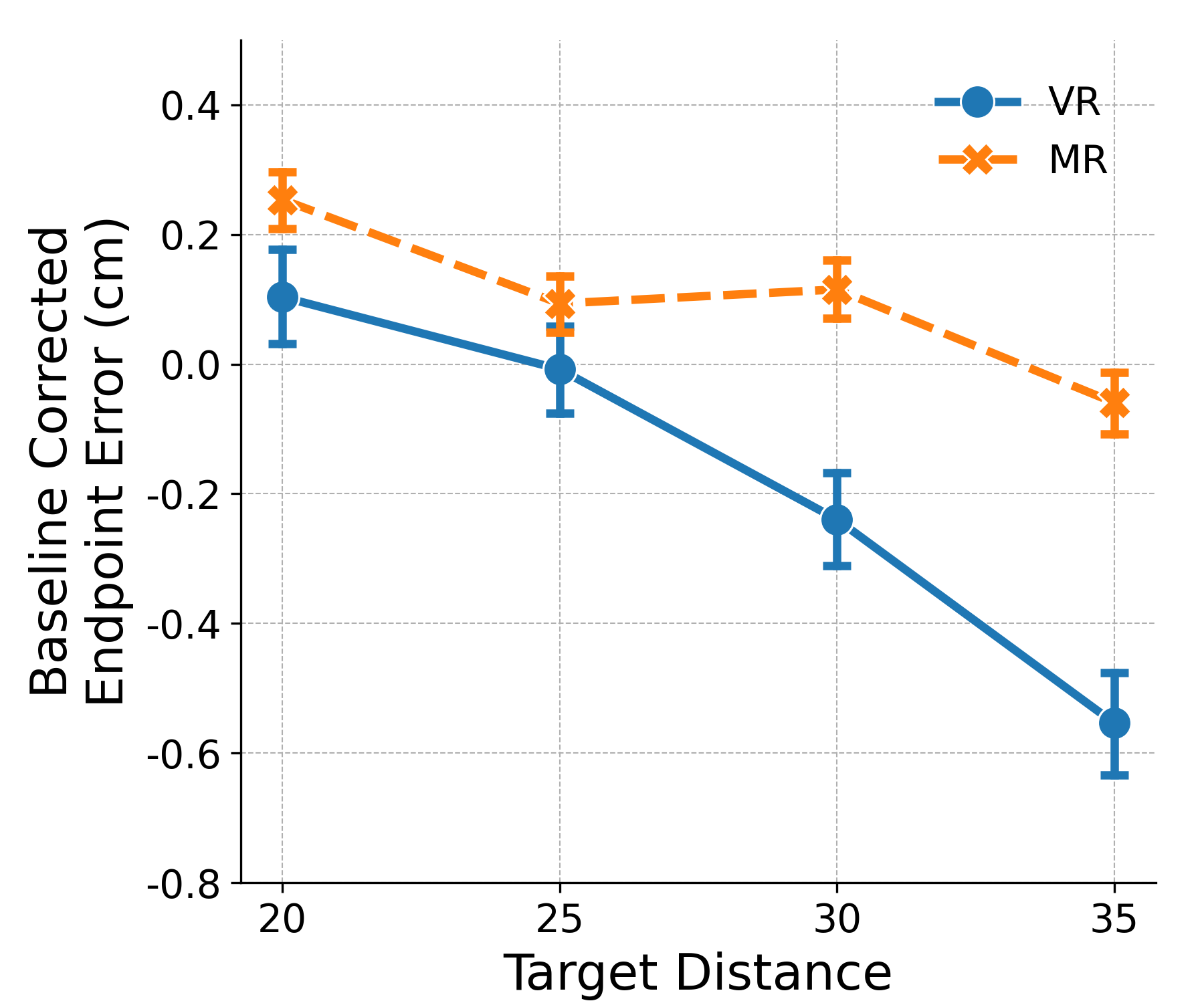}
  \caption{Baseline-corrected endpoint error ($EE_{bc}$) as a function of target distance (20--35~cm) during the XR phase. Values were first averaged across XR blocks within participants, then averaged within group. Negative values indicate undershooting relative to each participant's pre-intervention baseline. Reaching error became increasingly negative with target distance in VR, whereas the distance effect was shallower in MR. Error bars represent 95\% confidence intervals.}
  \label{fig:xr_vac}
\end{figure}

\subsection{UR Pre-Intervention Phase}

\subsubsection*{Endpoint Error}

To establish baseline accuracy and verify group equivalence prior to XR exposure, EE was analyzed using an LMM for the UR Pre-Intervention phase. The groups had similar overall baseline MT. Although the MR-VPT group exhibited a slightly steeper distance--MT relationship, the magnitude of this difference was small. The LMM revealed no significant main effect of  Modality, $F(1, 37.19) = 0.19$, $p = .669$, $\eta_p^2 = .005$. None of the interactions involving Modality were significant, including Modality $\times$ Block, $F(1, 37.06) = 0.78$, $p = .384$, and Modality $\times$ Target Distance, $F(1, 2254.36) = 0.01$, $p = .910$. Estimated marginal slopes likewise showed similar changes across blocks between VR and MR-VPT (VR: $-0.012$ cm/block, 95\% CI $[-0.026, 0.002]$; MR: $-0.003$ cm/block, 95\% CI $[-0.018, 0.012]$; contrast: $t(37) = 0.88$, $p = .384$). Comparable distance-dependent trends were also observed between the two modalities (VR: $-0.055$ cm/5 cm, 95\% CI $[-0.077, -0.034]$; MR: $-0.057$ cm/5 cm, 95\% CI $[-0.079, -0.035]$; contrast: $t(2254) = -0.11$, $p = .910$). Together, these findings confirm that the VR and MR-VPT groups were comparable prior to XR exposure, providing a common baseline for subsequent comparisons.

Performance remained stable across the Pre-Intervention phase (Figure~\ref{fig:ee_mt}a). Block was not significant, $F(1, 37.06) = 2.25$, $p = .142$, $\eta_p^2 = .06$, suggesting no systematic learning or performance drift across the Pre-Intervention blocks. Across both groups, participants exhibited a small overshoot (VR: $M = 0.38$~cm, $SD = 0.58$; MR: $M = 0.46$~cm, $SD = 0.47$), which diminished with increasing target distances (20~cm: $M = 0.51$~cm; 25~cm: $M = 0.42$~cm; 30~cm: $M = 0.41$~cm; 35~cm: $M = 0.33$~cm). Consistent with this pattern, Target Distance was the only significant predictor, $F(1, 2259.51) = 19.14$, $p < .001$, $\eta_p^2 = .008$. EE decreased by an estimated 0.056~cm for each 5~cm increase in target distance (95\% CI $[-0.072, -0.041]$). This distance effect remained stable across the Pre-Intervention phase, as indicated by the non-significant Block $\times$ Target Distance interaction, $F(1, 2258.71) = 0.32$, $p = .574$. Overall, the effect of Target Distance was consistent with the classic range effect observed in manual aiming, in which overshoot decreases as movement distance increases \cite{ElliottLee:1995:PR}. 

\subsubsection*{Movement Time}

To verify group equivalence in baseline movement timing prior to the XR exposure, MT was analyzed using the same LMM as EE. The VR and MR-VPT groups entered the XR phase with similar mean baseline MT, despite a small difference in distance scaling. The LMM revealed no significant main effect of Modality, $F(1, 36.99) = 1.97$, $p = .169$, $\eta_p^2 = .05$. The Modality $\times$ Block interaction was also not significant, $F(1, 36.99) = 2.21$, $p = .145$, indicating that the MT trajectories did not differ between groups across the Pre-Intervention phase (Figure~\ref{fig:ee_mt}d). 

Movement timing also remained stable across the Pre-Intervention phase. Block was not significant, $F(1, 36.99) = 0.39$, $p = .535$, $\eta_p^2 = .01$, indicating no systematic learning or performance drift across blocks. Across both groups, participants produced movements with a mean MT of 0.43~s.

Finally, MT was longer for farther targets, as indicated by a strong main effect of Target Distance, $F(1, 2250.70) = 422.71$, $p < .001$, $\eta_p^2 = .16$, with MT increasing by an estimated 0.034~s for every 5~cm increase in target distance (95\% CI $[0.032, 0.036]$). This distance effect remained stable across the Pre-Intervention phase, as indicated by the non-significant Block $\times$ Target Distance interaction, $F(1, 2250.65) = 0.02$, $p = .887$. The Modality $\times$ Target Distance interaction was significant, $F(1, 2250.30) = 14.67$, $p < .001$, $\eta_p^2 = .006$, reflecting a steeper distance--MT relationship for MR-VPT ($+0.037$ s/5 cm, 95\% CI $[0.035, 0.040]$) than for VR ($+0.030$ s/5 cm, 95\% CI $[0.028, 0.033]$; contrast: $t(2250) = 3.83$, $p < .001$). However, this difference was small (approximately 7~ms per 5~cm increase in target distance) and was therefore unlikely to represent a practically meaningful baseline difference.

\subsection{XR Phase}

\subsubsection*{Endpoint Error}

To characterize adaptation during the XR phase, baseline-corrected EE ($EE_{bc}$) was modeled using an exponential-to-asymptote NLMM across Block and Target Distance (Figure~\ref{fig:ee_mt}b). The analysis revealed three key differences between VR and MR: opposite initial endpoint biases, different adaptation rates, and different distance-dependent EE. Despite these differences, both groups converged to comparable adapted states by the end of the XR phase.

\smallskip

\noindent\textbf{Initial bias ($EE_0$).} 
The two modalities produced significantly different initial endpoint biases at the onset of the XR phase (LRT: $\chi^2(1) = 19.31$, $p < .001$). The MR-VPT group entered block 0 with a positive $EE_{bc}$ ($EE_0 = 0.43$ cm, 95\% CI $[0.20, 0.65]$), indicating an overshoot relative to the UR baseline, whereas the VR group showed an initial undershoot ($EE_0 = -0.31$ cm, 95\% CI $[-0.52, -0.10]$). These opposite initial biases indicate that the two XR modalities initially perturbed the mapping between perception and action in opposite directions.

\smallskip

\noindent\textbf{Adaptation rate ($\tau$).} 
The time constant differed significantly between modalities (LRT: $\chi^2(1) = 9.24$, $p < .01$). MR-VPT participants adapted rapidly ($\tau_{\mathrm{MR}} = 1.86$ blocks), reaching approximately 95\% of their respective adapted state within the first six blocks ($t_{95} \approx 5.6$ blocks). By contrast, VR participants adapted substantially more slowly ($\tau_{\mathrm{VR}} = 4.61$ blocks, $t_{95} \approx 13.8$ blocks), suggesting that VR approached its estimated asymptote only during the final portion of the XR phase.

\smallskip

\noindent\textbf{Adapted state ($A$).} 
Despite the different adaptation rates, the asymptotes did not differ significantly between modalities (LRT: $\chi^2(1) = 1.52$, $p = .218$). MR-VPT reached an asymptote of $A = 0.05$ cm (95\% CI $[-0.09, 0.18]$), whereas VR reached $A = -0.11$ cm (95\% CI $[-0.27, 0.04]$). Together, these results indicate that both groups approached states close to baseline-corrected accuracy.

\smallskip

\noindent\textbf{Distance effect.} 
Distance-dependent EE was stronger in VR than in MR-VPT and remained stable across the XR phase. Target Distance showed a significant main effect on $EE_{bc}$ ($F(1, 4690) = 299.93$, $p < .001$), with a shared slope of $\beta_{\mathrm{dist}} = -0.13$ cm/5 cm (95\% CI $[-0.17, -0.10]$), indicating increasing undershoot as target distance increased. This distance effect differed significantly between modalities (LRT: $\chi^2(1) = 48.32$, $p < .001$), where the distance slope was substantially steeper in VR ($\beta_{\mathrm{dist,VR}} = -0.19$ cm/5 cm, 95\% CI $[-0.26, -0.12]$) than in MR-VPT ($\beta_{\mathrm{dist,MR}} = -0.07$ cm/5 cm, 95\% CI $[-0.11, -0.03]$; Figure~\ref{fig:xr_vac}). In fact, the MR-VPT group's slope in the XR phase was comparable to that in the Pre-Intervention phase ($\beta_{\mathrm{dist}} = -0.056$, 95\% CI $[-0.072, -0.041]$), suggesting that the modality difference was driven primarily by the steeper VR distance scaling. The interaction between Block and Target Distance was not significant ($F(1, 4690) = 2.96$, $p = .086$), nor did it differ by Modality (LRT: $\chi^2(1) = 1.29$, $p = .255$), indicating that the distance-dependent error remained stable across the adaptation period.

\subsubsection*{Movement Time}

To characterize changes in movement timing during the XR phase, baseline-corrected MT ($MT_{bc}$; hereafter MT) was analyzed using a quadratic LMM. The analysis revealed different MT trajectories in VR and MR-VPT across practice. MT decreased across the XR phase in VR but remained relatively stable in MR. Despite these different adaptation trajectories, the two groups exhibited comparable MT at both the beginning and end of the XR phase (Figure~\ref{fig:ee_mt}e).

\smallskip

\noindent\textbf{Adaptation trajectory.}
There was no overall difference in MT between VR and MR, as the main effect of Modality was not significant, $F(1, 37.7) = 2.43$, $p = .128$. However, MT did change across the XR phase, as indicated by significant linear ($F(1, 37.2) = 7.23$, $p < .05$) and quadratic ($F(1, 4655.5) = 8.71$, $p < .01$) effects of Block. Importantly, the Modality $\times$ Block$^2$ interaction was significant, $F(1, 4655.5) = 53.60$, $p < .001$, $\eta_p^2 = .01$, whereas the Modality $\times$ Block interaction was not, $F(1, 37.2) = 0.69$, $p = .410$. Together, these findings indicate that the modality difference was expressed primarily in the curvature of the MT trajectory rather than in its overall linear trend.

Examining each modality separately, VR participants exhibited significantly shorter MT at the final XR block than at the first block, $t(36.7) = 2.59$, $p < .05$, $\Delta \approx 25$ ms, whereas MR-VPT participants showed no significant change across the XR phase, $t(37.3) = 1.27$, $p = .213$, $\Delta \approx 13$ ms. Despite these different within-modality changes, the two groups did not differ significantly at either the first XR block, $t(38.8) = -0.60$, $p = .553$, or the final XR block, $t(38.5) = 0.07$, $p = .946$. These simple-effects comparisons are consistent with the significant Modality $\times$ Block$^2$ interaction, indicating that the modalities differed primarily in how MT changed over practice rather than in overall MT.

\smallskip

\noindent\textbf{Distance effect.}
MT increased with target distance in both modalities, although this increase was more gradual in VR than in MR. Target Distance showed a strong main effect on MT, $F(1, 4653.7) = 2895.71$, $p < .001$, $\eta_p^2 = .38$, indicating progressively longer MT for farther targets. This distance effect also differed significantly between modalities (Modality $\times$ Distance: $F(1, 4653.7) = 8.61$, $p < .01$), indicating that MT increased less steeply with target distance in VR than in MR.

\subsection{UR Post-Intervention Phase}

\subsubsection*{Endpoint Error}

To evaluate aftereffects and de-adaptation following HMD removal, baseline-corrected EE ($EE_{bc}$; hereafter EE) was modeled using the same exponential-to-asymptote NLMM as in the XR phase. Unlike the XR phase, in which VR and MR-VPT produced opposite initial endpoint biases, both modalities exhibited immediate undershooting following HMD removal. The analysis further revealed that both groups rapidly de-adapted toward baseline, although only VR retained a persistent residual undershoot by the end of the Post-Intervention phase.

\smallskip

\noindent\textbf{Aftereffect ($EE_0$).} 
Both modalities showed significant negative EE at the start of the Post-Intervention phase, indicating immediate undershooting relative to their Pre-Intervention baseline following HMD removal. MR-VPT participants undershot by $EE_0 = -0.35$ cm (95\% CI $[-0.495, -0.202]$), confirming a significant aftereffect. The VR group exhibited a numerically larger aftereffect ($EE_0 = -0.51$ cm, 95\% CI $[-0.66, -0.37]$), although the difference between modalities did not reach significance (LRT: $\chi^2(1) = 2.72$, $p = .099$).

\smallskip

\noindent\textbf{De-adaptation rate ($\tau$).}
The de-adaptation rate did not differ significantly between modalities (LRT: $\chi^2(1) = 0.98$, $p = .323$). MR-VPT exhibited a washout time constant of $\tau = 2.12$ blocks ($t_{95} \approx 6.4$ blocks), whereas VR exhibited $\tau = 1.04$ blocks ($t_{95} \approx 3.1$ blocks). Thus, both modalities rapidly de-adapted toward baseline during the early Post-Intervention blocks.

\smallskip

\noindent\textbf{Final state ($A$).} 
Despite similar de-adaptation rates, the estimated final state of washout differed significantly between modalities (LRT: $\chi^2(1) = 6.18$, $p < .05$). MR-VPT reached an asymptote near baseline with the confidence interval including zero ($A = -0.106$ cm, 95\% CI $[-0.250, 0.038]$). By contrast, VR converged to a negative asymptote with the confidence interval remaining below zero ($A = -0.354$ cm, 95\% CI $[-0.494, -0.214]$). These findings indicate that MR-VPT participants returned to a level not significantly different from its UR baseline, whereas VR participants retained a persistent residual undershoot and did not fully de-adapt during the approximately 10--15 minute Post-Intervention phase.

\smallskip

\noindent\textbf{Distance effect.} 
Distance-dependent EE persisted immediately following HMD removal but gradually diminished as participants de-adapted and recovered the pre-exposure perception-action mapping. Target Distance showed a significant main effect ($\beta_{\mathrm{dist}} = -0.105$ cm/5 cm, 95\% CI $[-0.126, -0.085]$, $t(4606) = -10.08$, $p < .001$), indicating greater undershooting for farther targets. Importantly, the Block $\times$ Target Distance interaction was significantly positive ($\beta_{\mathrm{block} \times \mathrm{dist}} = 0.003$ cm/block/5 cm, 95\% CI $[0.001, 0.006]$, $p < .01$), indicating that this distance-dependent undershoot progressively diminished throughout the Post-Intervention phase.

The Modality $\times$ Distance interaction did not reach significance ($F(1,4613)=3.76$, $p=.053$). Estimated distance slopes were $-0.093$ cm/5 cm for MR-VPT (95\% CI $[-0.116,-0.070]$) and $-0.066$ cm/5 cm for VR (95\% CI $[-0.089,-0.044]$), with the pairwise contrast likewise not reaching significance ($t(4623)=-1.61$, $p=.107$), indicating that the recovery of distance-dependent error was comparable across modalities.

\subsubsection*{Movement Time}

To characterize changes in movement timing following HMD removal, baseline-corrected MT ($MT_{bc}$; hereafter MT) was analyzed using a quadratic LMM. Unlike the XR phase, the Post-Intervention phase showed no evidence that MT recovery differed between modalities. Instead, both VR and MR-VPT exhibited comparable MT trajectories throughout the washout period.

\smallskip

\noindent\textbf{De-adaptation trajectory.}
MT changed across the Post-Intervention phase, following a shared nonlinear trajectory across modalities. The quadratic Block$^2$ term was significant, $F(1, 4570.0) = 14.01$, $p < .001$, $\eta_p^2 = .003$, whereas the Modality $\times$ Block$^2$ interaction was not significant, $F(1, 4570.0) = 3.06$, $p = .080$, $\eta_p^2 < .001$. These findings suggest that the curvature of the recovery trajectory did not differ reliably between VR and MR.

No significant main effect of Modality was observed, $F(1, 37.4) = 0.00$, $p = .985$, indicating no overall difference in MT between VR and MR-VPT during the Post-Intervention phase. Examining each modality separately, neither MR-VPT ($t(37.1) = -1.42$, $p = .166$) nor VR ($t(36.9) = -0.93$, $p = .358$) showed a significant change in MT between the first and final Post-Intervention blocks. Likewise, the two groups did not differ significantly at either the first Post-Intervention block ($t(38.5) = 0.26$, $p = .793$) or the final block ($t(37.8) = 0.42$, $p = .677$), indicating that MT recovery followed comparable trajectories in both modalities.

\smallskip

\noindent\textbf{Distance effect.}
MT was longer for farther targets, as shown in the significant effect of Target Distance, $F(1, 4569.7) = 2990.04$, $p < .001$, $\eta_p^2 = .40$. This distance effect also differed significantly between modalities (Modality $\times$ Distance: $F(1, 4569.7) = 51.12$, $p < .001$, $\eta_p^2 = .01$), with a steeper distance--MT slope in MR-VPT ($+0.037$ s/5 cm) than in VR ($+0.028$ s/5 cm; $t(4572) = 7.11$, $p < .001$). Because the same pattern was observed during both the Pre-Intervention and XR phases, this difference likely reflects a stable baseline characteristic rather than a consequence of XR exposure. The Block $\times$ Distance interaction was not significant, $F(1, 4570.6) = 0.73$, $p = .393$, indicating that the relationship between MT and target distance remained stable throughout the Post-Intervention phase.

\section{Discussion}

The current study involved comparisons of visually guided manual pointing in VR and MR-VPT using the same HMD and comparable experimental setup to determine whether shared hardware yields comparable perceptuomotor behavior across XR modalities. The findings indicate that it does not. Although both modalities were delivered through the same HMD with opaque displays, and therefore shared important display-level characteristics, they produced different initial biases, different adaptation dynamics, different distance-dependent EE scaling, and different post-exposure behavior despite converging to similar adapted states during the XR phase. Together, these findings suggest that perceptuomotor behavior cannot be assumed to generalize across VR and MR-VPT simply because they are delivered on the same hardware platform. The following discussion evaluates these findings in relation to the two competing perspectives introduced earlier, that is, whether perceptuomotor adaptation is governed primarily by shared display-level perturbations or by the broader action-relevant visual information available within each interaction modality.

\smallskip

\noindent\textbf{$H_1$: Different initial biases.} 
The first hypothesis was that VR and MR-VPT would produce different initial biases despite being delivered on the same HMD. This hypothesis was supported. At the onset of the XR phase, the two modalities showed significantly different EE: participants in MR-VPT began with a slight overshoot relative to the UR baseline, whereas participants in VR began with an initial undershoot. This qualitative pattern mirrors results from an earlier study that showed undershooting in VR and overshooting in MR-OST \cite{WangSouthwickRobinson:2024:SR}. In this earlier work, the opposite-signed biases were interpreted in part through a geometrical account of VAC, in which device-specific vergence-accommodation characteristics could produce systematic differences in pointing bias \cite{WangSouthwickRobinson:2024:VR, WangSouthwickRobinson:2024:SR}. Because VR and MR-OST necessarily relied on different display hardware, however, it remained unclear whether these opposite biases reflected display optics, interaction modality, or both.

By delivering both modalities on the same HMD (Meta Quest 3), the present study controlled many display-level constraints that differed in the earlier VR/MR-OST comparison. The opposite-signed initial biases therefore suggest that modality-specific factors beyond shared hardware contributed to the initial perturbation. A key difference between VR and MR-VPT is the visual context in which the action is embedded. In VR, the entire visual scene was synthetic, whereas in MR-VPT, participants acted within a camera-mediated view of the physical world that preserved scene structure and visual access to the body. Moreover, the camera-mediated view is not a raw camera feed but is instead synthesized from head-mounted camera streams and depth estimates. These views are then depth-warped from the cameras' viewpoints into per-eye views aligned with the user's eye positions, compensating for camera--eye offsets and head motion \cite{KuoPennerMoczydlowski:2023:SIGGRAPH}. Such processing can introduce local geometric distortions, particularly around depth discontinuities \cite{KuoPennerMoczydlowski:2023:SIGGRAPH, MetaQuest_v66_2024}, and may therefore produce additional perceptuomotor perturbations. Together, these differences suggest that VR and MR-VPT do not simply differ in scene appearance; they provide different action-relevant information for guiding movement. The opposite initial biases observed here may therefore reflect the combined influence of preserved physical context, body visibility, and passthrough-specific spatial distortions rather than display characteristics alone. 

The present findings extend prior same-headset comparisons by showing that modality differences are evident not only in perceptual judgment tasks \cite{WestermeierBruebachWienrich:2024:TVCG}, but also during goal-directed manual interactions. These findings therefore favor the view that initial perceptuomotor perturbations depend not only on shared display-level constraints but also on the broader action-relevant visual information available within each interaction modality.

\smallskip

\noindent\textbf{$H_2$: Different adaptation processes.}
The second hypothesis was that adaptation would unfold differently across modalities, with performance in MR-VPT stabilized more rapidly than in VR. This hypothesis was supported. Although performance in both modalities ultimately converged to comparable adapted states, participants in the MR-VPT group reached that state substantially faster than those in VR, indicating that the main difference between the two modalities lay in the adaptation process rather than in steady-state performance. This pattern is consistent with prior work showing that users can adapt to perceptuomotor perturbations in immersive displays \cite{KohmBabuPagano:2022:TVCG}, while extending that work by showing that the rate of adaptation can differ across XR modalities even when display hardware is held constant.

The comparable adapted states indicate that participants in both modalities ultimately recalibrated to the perturbed perception-action mapping and regained stable perceptuomotor performance through practice. However, the different adaptation rates suggest that VR and MR-VPT differed less in the stable performance ultimately achieved than in the recalibration demands required to achieve it. Because both modalities shared the same headset and fixed-focal display, these differences are more likely to reflect the broader action-relevant visual information available during interaction. MR-VPT preserved the physical table, physical targets, and the participant's own body within a camera-mediated view of the room, whereas VR required action toward virtual targets with feedback from a bodyless virtual hand in a synthetic scene. These differences may influence how readily participants relate visual feedback to their physical hand and the surrounding spatial layout \cite{ScheidtCondittSecco:2005:JNP}. Consequently, the faster stabilization observed in MR-VPT is best interpreted as a property of the implemented interaction environment rather than as evidence for a single isolated mechanism. Practically, these findings suggest that studies relying on brief XR exposures may primarily capture transient recalibration dynamics rather than steady-state performance. Although VR initially required more extensive recalibration than MR-VPT, sufficient practice allowed both modalities to reach comparable levels of pointing accuracy.

Finally, MT showed a qualitatively similar but weaker pattern than EE. Although participants in the VR condition exhibited a modest reduction in MT across the XR phase whereas those in MR-VPT did not, the two modalities did not differ at either the beginning or the end of XR exposure. Thus, the modality effect was again expressed primarily in the adaptation trajectory rather than in steady-state performance. Unlike EE, however, MT differences were comparatively small and were confined mainly to the early stages of adaptation. This pattern suggests that the main modality differences observed in the present study reflect recalibration of the perception-action mapping rather than a general slowing or speeding of movement execution. Consistent with this interpretation, the clearest modality differences emerged in spatial accuracy (EE), whereas movement timing remained comparatively similar throughout adaptation.

\smallskip

\noindent\textbf{$H_3$: Stronger distance-dependent endpoint error scaling in VR.}
The third hypothesis was that distance-dependent EE would be stronger in VR than in MR-VPT. This hypothesis was supported. During XR, baseline-corrected EE became increasingly negative with greater target distance in both modalities, but the distance dependence was much steeper in VR whereas MR-VPT closely approximated the slope observed during UR Pre-Intervention. Given that the UR Pre-Intervention slope is consistent with the classic range effect \cite{ElliottLee:1995:PR}, MR-VPT's distance dependence likely reflects this baseline tendency. In contrast, the much steeper distance scaling in VR indicates an additional depth-related error pattern during VR exposure.

Because VR and MR-VPT were delivered on the same HMD and shared the same display-level constraints, the stronger distance-dependent error observed in VR cannot be attributed solely to differences in display hardware. Instead, the results indicate that the implemented VR and MR-VPT interaction environments produced different depth-dependent patterns of perceptuomotor behavior. VR rendered the scene, targets, and virtual hand within a synthetic coordinate frame, whereas MR-VPT reconstructed the physical scene through camera-based passthrough while preserving the physical hand and targets. The present design does not distinguish which specific modality-level differences gave rise to the observed distance scaling, but it demonstrates that the overall interaction environment influences depth-dependent pointing behavior even when display hardware is held constant.

One possible account of this distance-dependent behavior is the geometrical model of VAC-related depth compression proposed in previous work \cite{WangPrenevostTarun:2026:Displays, WangSouthwickRobinson:2024:VR}. This model hypothesizes that VAC produces an inward vergence offset, perturbing the binocular viewing geometry and resulting in systematic distance-dependent undershooting during targeted reaching. To evaluate whether the present results were quantitatively compatible with this account, the supplementary appendix fits the XR data to the geometric VAC model after subtracting each participant's UR baseline pattern across target distances to estimate the vergence offset. The fitted model estimated a larger vergence offset in VR ($0.35^\circ$, 95\% participant-bootstrap CI $[0.25, 0.45]$) than in MR-VPT ($0.06^\circ$, 95\% CI $[-0.02, 0.15]$), with a significant modality difference ($\Delta\beta = 0.29^\circ$, 95\% CI $[0.16, 0.43]$, bootstrap $p = .002$). Thus, the observed distance scaling is quantitatively compatible with a VAC-related depth-scaling distortion. However, because the present study manipulated XR modality rather than VAC itself, the current design does not establish VAC as the causal source of the modality difference. Importantly, the distance dependence remained stable across XR exposure. Thus, although practice reduced the overall error offset, it did not eliminate the systematic tendency for farther targets to elicit greater undershoot, particularly in VR. Together, these findings suggest that perceptuomotor adaptation can reduce overall pointing error while leaving systematic distortions in distance scaling largely intact.

\smallskip

\noindent\textbf{$H_4$: More persistent aftereffects in VR.}
The fourth hypothesis was that both modalities would produce aftereffects when participants return to UR, and that participants who experienced VR would show larger and more persistent residual error than those in MR-VPT. This hypothesis was partially supported. Immediately upon returning to UR, participants in both groups exhibited a significant undershoot, indicating that exposure to both XR modalities altered subsequent performance in UR. Although the initial aftereffect was numerically larger in VR, this difference did not reach significance at Post-Intervention onset. Thus, the prediction of a larger initial VR aftereffect was not supported.

A notable pattern was that the MR-VPT group's Post-Intervention aftereffect was not aligned with the direction of its early XR perturbation. During the XR phase, MR-VPT began with an overshoot that rapidly returned toward baseline, whereas its Post-Intervention response showed undershooting similar in direction to VR, which also undershot during XR. This dissociation suggests that the initial XR perturbation in MR-VPT and the UR aftereffect may not reflect the same underlying process. In classic sensorimotor adaptation paradigms (e.g., visuomotor rotations), aftereffects during washout typically reflect persistence of the adapted motor command and therefore occur in the direction of that learned command (and opposite to the previously imposed perturbation) \cite{Krakauer:2009:PMC, KrakauerGhezGhilardi:2005:JN, KrakauerPineGhilardi:2000:JN}. Applied to depth compression, accurate performance during XR would require reaching farther than the distorted percept (i.e., an overshoot-like compensatory command). If this compensatory command persisted after HMD removal, the initial Post-Intervention response would be expected to overshoot rather than undershoot. The observed initial undershoot in both modalities is therefore not readily explained as a straightforward motor aftereffect of compensating for XR depth compression alone. One possible contributor is short-term adaptation in vergence-accommodation coupling following prolonged fixed-focal HMD viewing \cite{NeveuRoumesPhilippe:2016:IOVS, PaulusStraubeEggert:2017:JN, YegoGilsonBaraas:2025:Displays, YegoGilsonBaraas:2025:IOVS}. Such adaptation could contribute to undershooting after HMD removal in both groups, but it is unlikely to fully explain the observed modality differences because distance-dependent undershoot during XR differed substantially between VR and MR-VPT. More work is therefore needed to separate post-exposure oculomotor effects from modality-specific recalibration of the perception-action mapping.

While the initial aftereffects were similar in direction across modalities, the later stages of de-adaptation revealed a different pattern, where performance in VR and MR-VPT converged to different final states during washout. Specifically, participants who experienced MR-VPT returned to a level not reliably different from UR baseline, whereas those who experienced VR retained a significant residual undershoot even at the end of the Post-Intervention phase. This dissociation, that is, similar initial aftereffects and de-adaptation dynamics but different residual bias, is consistent with multi-rate state-space accounts in which fast processes drive rapid early recovery while slower, more strongly retained processes determine the residual bias remaining within a finite washout period \cite{SmithGhazizadehShadmehr:2006:PLoSB}. One possibility is that the richer action-relevant visual information available in MR-VPT facilitated more effective separation between XR-specific and UR-specific perceptuomotor mappings, allowing rapid reinstatement of the UR calibration after HMD removal. By contrast, VR may have promoted a more strongly retained recalibration because the bodyless synthetic environment provided fewer stable real-world anchors for distinguishing between the two contexts. Although speculative, this interpretation is consistent with evidence that contextual cues facilitate context-dependent motor memory and switching between learned sensorimotor mappings \cite{AvrahamTaylorBreska:2022:eLife}.

\subsection{Limitations and Future Directions}

Several limitations should be acknowledged. First, the study compared two realistic XR implementations rather than independently manipulating each factor that differed between them. Thus, Modality in this experiment bundled scene construction, target presentation, body representation, passthrough reconstruction, and display-mediated viewing. This design supports conclusions about the VR and MR-VPT implementations on the Quest 3 HMD, but it does not identify which individual factor was responsible for each behavioral difference.

A second limitation is that the two interaction environments differed simultaneously in several sources of action-relevant visual information. While participants had visual access to their physical body in MR-VPT, those in VR could only see and interact via an opaque, disembodied hand (Figure~\ref{fig:study-overview}). Prior work suggests that precise VR interaction is sensitive to how the user's hands are rendered. For example, Turkmen et al.~\cite{TurkmenVoisardKerstenOertel:2025:ISMAR} showed that dynamically adjusting hand visibility across interaction phases, i.e., keeping the hand visible during the approach phase while reducing its visibility during the targeting phase, improved both accuracy and usability compared with a consistently opaque hand. More broadly, in immersive VR, using video-passthrough to enable interactions via the user's physical body has been shown to enhance presence and body ownership while maintaining task performance and cognitive load \cite{WaldowKleinbeckFuhrmann:2025:TVCG}. Although effects on performance may be task-dependent, increased embodiment and body ownership have also been linked to improved motor accuracy and reduced completion time in 3D tracking \cite{OdermattBuetlerWenk:2021:FN}. Therefore, differences in body representation and embodiment between the VR condition without a full-body avatar and the MR-VPT condition may have contributed to the observed modality differences. Future work should systematically manipulate body representation while holding the interaction environment constant, for example by comparing a virtual hand, a full-body avatar, and video-passthrough embodiment within the same VR environment, to determine how body visibility and embodiment influence initial perturbation, adaptation dynamics, and post-exposure transfer.

Similarly, the MR-VPT condition also did not fully exploit the defining feature of mixed reality, namely displaying virtual objects that are spatially co-registered with the physical environment. Participants primarily viewed the real world through passthrough while performing the task, rather than interacting with virtual objects anchored to physical surfaces, such as virtual targets presented on the physical table. Future work should systematically manipulate target cueing, physical anchoring, environmental context, visual realism, and passthrough reconstruction to determine how these sources of action-relevant visual information contribute to initial perturbation, adaptation dynamics, distance-dependent error, and post-exposure transfer.

Finally, the geometric VAC model presented in the supplementary appendix should be interpreted as a quantitative consistency check rather than as a causal explanation of the observed modality differences. Although the estimated vergence offsets indicated that the observed distance scaling was quantitatively compatible with previous geometrical accounts of VAC, the present design did not independently manipulate VAC or other display-level constraints. Future work should therefore combine controlled manipulations of display-level perturbations (e.g., VAC, motion-to-photon latency) with systematic manipulations of action-relevant visual information to determine how these factors independently and jointly shape perceptuomotor adaptation and transfer across XR modalities.

\section{Conclusion}

In summary, the present study showed that VR and MR-VPT, even when delivered on the same HMD, do not produce equivalent perceptuomotor behavior. Although both modalities ultimately converged to comparable adapted states during XR exposure, they differed systematically in their initial perturbations, adaptation dynamics, distance-dependent endpoint error, and post-exposure behavior. These findings demonstrate that shared display hardware alone is insufficient to predict perceptuomotor adaptation in XR. Instead, adaptation depends on the broader interaction environment in which visually guided action is embedded. More broadly, the results suggest that display-level constraints and action-relevant visual information should be considered together when evaluating, designing, and comparing immersive interaction techniques across XR modalities.

\appendix

\section{Geometric VAC Model Analysis}
\label{app:vac-model}

\setcounter{equation}{0}
\setcounter{table}{0}
\setcounter{figure}{0}
\renewcommand{\theequation}{\thesection\arabic{equation}}
\renewcommand{\thetable}{\thesection\arabic{table}}
\renewcommand{\thefigure}{\thesection\arabic{figure}}
\renewcommand{\theHequation}{\thesection\arabic{equation}}
\renewcommand{\theHtable}{\thesection\arabic{table}}
\renewcommand{\theHfigure}{\thesection\arabic{figure}}

This appendix provides a quantitative consistency check for the distance-scaling result reported in the main text. The goal was not to demonstrate that the vergence-accommodation conflict (VAC) caused the modality difference, but to test whether the residual distance-dependent endpoint error (EE) after baseline correction followed the distance-scaling pattern predicted by a VAC model.

\subsection{Model Rationale}

The main analysis showed that baseline-corrected EE in the XR phase became increasingly negative with target distance, with a steeper distance slope in VR ($-0.19$~cm/5~cm) than in MR-VPT ($-0.07$~cm/5~cm). Because the UR Pre-Intervention phase already showed a small distance slope consistent with the classical range effect, fitting a geometric VAC model to raw EE would incorrectly attribute the baseline range effect to VAC. We therefore computed a distance-specific baseline correction: for each participant, mean UR Pre-Intervention EE was subtracted separately at each target distance before fitting the geometric model to XR and UR Post-Intervention cells.

Following Wang et al.~\cite{WangPrenevostTarun:2026:Displays, WangSouthwickRobinson:2024:VR}, VAC was modeled as a constant vergence offset $\beta$ added to the physical vergence angle. For target depth $d$, the physical vergence angle $\theta_{\mathrm{phys}}(d)$ becomes

\begin{equation}
  \theta_{\mathrm{percept}}(d) = \theta_{\mathrm{phys}}(d) + \beta .
\end{equation}

\noindent 
Under this formulation, $\beta$ indexes the magnitude of the hypothesized vergence offset. The resulting function predicts how VAC would reshape EE across target distances, allowing the observed distance-scaling pattern to be compared with the model prediction. The resulting stereoscopic viewing distance is
\begin{equation}
  D_{\mathrm{stereo}}(d;\beta) =
  \frac{\mathrm{IPD}/2}{\tan(\theta_{\mathrm{percept}}(d)/2)} .
\end{equation}
Because the distance-independent component of the prediction is not identifiable separately from general adaptation bias, the fitted VAC term used the centered distance shape
\begin{equation}
  s(d;\beta) = f(d;\beta) -
  \frac{1}{4}\sum_{d' \in \{20,25,30,35\}} f(d';\beta),
\end{equation}
where $f(d;\beta)$ is the EE predicted from the VAC geometry at distance $d$ before removing its mean across target distances. The intercept across blocks was modeled separately with the same exponential-to-asymptote form used in the main analysis, so $\beta$ was identified from distance-dependent shape rather than overall bias.

\subsection{Model Comparison and Offset Estimates}

Three nested models were fit separately by modality:
\begin{itemize}[itemsep=0pt, topsep=4pt]
    \item[$M_0$] an intercept-only model with adaptation but no VAC-predicted distance-scaling term;
    \item[$M_1$] a constant vergence offset VAC model;
    \item[$M_2$] a dynamic vergence offset VAC model allowing $\beta$ to vary across blocks.
\end{itemize}

The constant-offset model was preferred by BIC in both modalities (Table~\ref{tab:vac-model-comparison}). Although the dynamic model produced slightly higher $R^2$ and lower AIC, BIC consistently favored the simpler constant-offset model in both modalities. Adding the centered VAC-predicted distance-scaling term improved fit most strongly in VR, where $R^2$ increased from $0.28$ to $0.65$. In MR-VPT the improvement was smaller ($0.76$ to $0.81$), consistent with the main text's finding that MR-VPT distance scaling closely approximated the baseline range effect.

\begin{table}[tb]
\centering
\small
\caption{Goodness of fit for nested VAC models. Lower AIC/BIC is better; the preferred model by BIC is in bold.}
\label{tab:vac-model-comparison}
\begin{tabular}{llccc}
\toprule
Modality & Model & $R^2$ & AIC & BIC \\
\midrule
MR-VPT & Intercept-only ($M_0$) & 0.760 & $-259.4$ & $-243.0$ \\
MR-VPT & \textbf{Constant VAC ($M_1$)} & \textbf{0.811} & $\mathbf{-285.3}$ & $\mathbf{-263.7}$ \\
MR-VPT & Dynamic VAC ($M_2$) & 0.831 & $-290.6$ & $-259.1$ \\
\midrule
VR & Intercept-only ($M_0$) & 0.280 & $-72.3$ & $-55.9$ \\
VR & \textbf{Constant VAC ($M_1$)} & \textbf{0.650} & $\mathbf{-160.0}$ & $\mathbf{-138.4}$ \\
VR & Dynamic VAC ($M_2$) & 0.660 & $-154.4$ & $-122.9$ \\
\bottomrule
\end{tabular}
\end{table}

\begin{figure}[tb]
\centering
\includegraphics[width=.85\columnwidth]{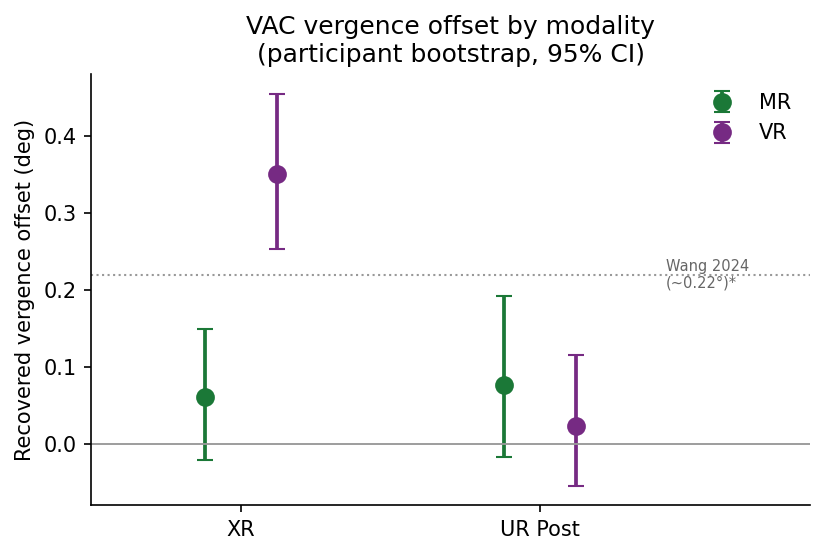}
\caption{Estimated vergence offset by modality and phase from the constant-offset VAC model. Points and intervals show participant-bootstrap means and 95\% confidence intervals.}
\label{fig:vac-offset}
\end{figure}

Participant-bootstrap inference showed that the estimated constant vergence offset during the XR phase was larger in VR ($\beta = 0.35^\circ$, 95\% CI $[0.25, 0.45]$) than in MR-VPT ($\beta = 0.06^\circ$, 95\% CI $[-0.02, 0.15]$; Figure~\ref{fig:vac-offset}). The modality difference was reliable ($\Delta\beta = 0.29^\circ$, 95\% CI $[0.16, 0.43]$, bootstrap $p = .002$). In the UR Post-Intervention phase, offsets were small and did not reliably differ between modalities (VR: $0.03^\circ$, 95\% CI $[-0.05, 0.12]$; MR-VPT: $0.07^\circ$, 95\% CI $[-0.02, 0.19]$; $\Delta\beta = -0.05^\circ$, 95\% CI $[-0.19, 0.08]$).

\subsection{Interpretive Boundaries}

This analysis supports the main-text claim that the stronger VR distance-scaling pattern is quantitatively compatible with VAC-related depth compression. Several constraints limit the interpretation. First, the fitted offset uses a centered shape term, so its absolute magnitude is not directly comparable to offsets from uncentered participant-level fits with free IPD. Second, IPD was fixed to a constant value rather than measured per participant. Third, the geometric model was estimated from only four target distances, so the offset estimate is stabilized by pooling across blocks rather than by dense sampling of the distance function. Finally, the experimental design did not independently manipulate VAC, body representation, target cueing, scene context, or passthrough reconstruction. Accordingly, the model should be interpreted as evidence of geometric compatibility rather than as evidence that VAC was the primary cause of the modality difference.


\acknowledgments{
We acknowledge the support of the Natural Sciences and Engineering Research Council of Canada (NSERC).

AI tools were used for limited implementation-related consultation for statistical analysis, including code refinement, debugging, and clarification of statistical programming steps (Claude), and for language editing and wording refinement (ChatGPT). All analytic decisions, model specification, result verification, interpretation, and conclusions were performed by the authors.
}

\bibliographystyle{abbrv-doi}

\bibliography{quest_mr_vr_preprint_v1}

\end{document}